\documentclass[draft,twocolumn,prl]{revtex4}

\usepackage{amsmath,amsfonts,amssymb}

\newtheorem{theorem}{Theorem}
\newtheorem{corollary}[theorem]{Corollary}
\newtheorem{lemma}[theorem]{Lemma}
\newtheorem{proposition}[theorem]{Proposition}
\newcommand{\qed}{\rule{7pt}{7pt}}
\newenvironment{proof}{\noindent{\bf Proof}\hspace*{1ex}}{\qed\bigskip}

\DeclareMathOperator{\tr}{Tr}
\DeclareMathOperator{\qu}{qubit}
\DeclareMathOperator{\cb}{cbit}
\DeclareMathOperator{\cc}{coherent\ bit}
\DeclareMathOperator{\eb}{ebit}
\DeclareMathOperator{\qus}{qubits}
\DeclareMathOperator{\cbs}{cbits}

\DeclareMathOperator{\coh}{cobit}
\DeclareMathOperator{\cohs}{cobits}
\DeclareMathOperator{\ebs}{ebits}
\DeclareMathOperator{\cat}{(c)}
\DeclareMathOperator{\asy}{(a)}

\def\be{\begin{equation}}
\def\ee{\end{equation}}
\def\bea{\begin{eqnarray}}
\def\eea{\end{eqnarray}}
\def\ben{\begin{eqnarray*}}
\def\een{\end{eqnarray*}}
\def\non{\nonumber}
\def\l{\left}
\def\r{\right}

\newcommand{\bra}[1]{\mbox{$\left\langle #1 \right|$}}
\newcommand{\ket}[1]{\mbox{$\left| #1 \right\rangle$}}

\newcommand{\eq}[1]{Eq.~(\ref{eq:#1})}

\def\ra{\rightarrow}
\def\la{\leftarrow}
\def\cE{{\cal E}}
\def\UA{U_{\bf A}}
\def\C{\mathbb{C}}
\def\cC{{\cal C}}
\def\cO{{\cal O}}
\def\cS{{\cal S}}
\def\cnot{\textsc{cnot}}

\begin{document}
\title{Coherent Communication of Classical Messages}
\author{Aram Harrow}
\affiliation{MIT Physics, 77 Massachusetts Ave, Cambridge, MA 02139}
\email{aram@mit.edu}
\date{\today}
\begin{abstract}
We define {\em coherent communication} in terms of a simple primitive,
show it is equivalent to the ability to send a classical message with
a unitary or isometric operation, and use it to relate other resources
in quantum information theory.  Using coherent communication, we are
able to generalize super-dense coding to prepare arbitrary quantum
states instead of only classical messages.  We also derive
single-letter formulae for the classical and quantum capacities of a
bipartite unitary gate assisted by an arbitrary fixed amount of entanglement
per use. 
\end{abstract}
\maketitle

\noindent{\bf Beyond qubits and cbits}

The basic units of communication in quantum information theory are
usually taken to be qubits and cbits, meaning one use of a noiseless
quantum or classical channel respectively.  Qubits and cbits can be
converted to each other and to and from other quantum communication
resources, but often only irreversibly.  For example, one qubit can
yield at most one cbit, but to obtain a qubit from cbits requires the
additional resource of entanglement.  By introducing a new resource,
intermediate in power between a qubit and a cbit, we will show that
many resource transformations can be modified to be reversible and
thus more efficient.

If $\{\ket{x}\}_{x=0,1}$ is a basis for $\C^2$, then a qubit channel
can be described as the isometry $\ket{x}_A\ra\ket{x}_B$ and a cbit
can be written as $\ket{x}_A\ra\ket{x}_B\ket{x}_E$.  Here $A$ is the
sender Alice, $B$ is the receiver Bob and $E$ denotes an inaccessible
environment, sometimes personified as a malicious Eve.  Tracing out
Eve yields the traditional definition of a cbit channel with the basis
$\{\ket{x}\}$ considered the computational basis.

Define a {\em coherent bit} (or ``cobit'') of communication as the
ability to perform the map $\ket{x}_A\ra \ket{x}_A\ket{x}_B$.  Since
Alice is free to copy or destroy her channel input, $1 \qu \geqslant 1
\coh\geqslant 1 \cb$, where $X\geqslant Y$ means that the resource $X$
can be used to simulate the resource $Y$.  We will also write
$X\geqslant Y\;\cat$ when $X+Z\geqslant Y+Z$ for some resource $Z$
used as a {\em catalyst} and $X \geqslant Y\;\asy$ when the conversion
is {\em asymptotic}, by which we mean that $X^{\otimes n}\agt
Y^{\otimes \approx n}$ for $n$ sufficiently large.

In Prop.~\ref{prop:unitary-cc}, we will see that coherent bits come
from any method for sending bits using a coherent procedure (a unitary
or isometry on the joint Alice-Bob Hilbert space); hence their name.
From their definitions, we see that cobits can be thought of as cbits
where Alice controls the environment, making them like classical
channels with quantum feedback.  Further connections between cobits
and cbits will be seen later in this Letter, where we show that
irreversible resource transformations are often equivalent to
performing ``1 cobit $\geqslant$ 1 cbit,'' and in \cite{family} where
several communication protocols are ``made
coherent'' with the effect of replacing cbits with cobits.

In the first half of this paper we will describe how to obtain
coherent bits and then how to use them, allowing us to exactly
describe the power of cobits in terms of conventional
resources.  The purpose of this paper is thus not to define a new
incomparable quantum resource, but rather to introduce a technique for
relating and composing communication protocols.  Then we will apply coherent
communication to remote state preparation\cite{RSP}.  This
will allow us to generalize super-dense coding and determine the
classical and quantum capacities of unitary gates with an arbitrary
amount of entanglement assistance.\\

\noindent{\bf Sources of coherent communication}

Qubits and cbits arise naturally from noiseless and dephasing channels
respectively, and can be obtained from any noisy channel by
appropriate coding \cite{HSW,Shor,Dev,Lloyd}.  Similarly, we will show both a
natural primitive yielding coherent bits and a coding theorem that
can generate coherent bits from a broad class of unitary operations.

The simplest way to send a coherent message is by modifying
super-dense coding\cite{SDC}.  In \cite{SDC}, Alice and Bob begin with
the state $\ket{\Phi_2}=\frac{1}{\sqrt{2}}\sum_{x=0}^1
\ket{x}_A\ket{x}_B$, which we call an {\em ebit}.  Alice encodes a
two-bit message $a_1a_2$ by applying $Z^{a_1}X^{a_2}$ to her half of
$\ket{\Phi_2}$ and then sending it to Bob, who decodes by applying
$(H\otimes I)\textsc{cnot}$ to the state, obtaining
$$(H\otimes I)\textsc{cnot}(Z^{a_1}X^{a_2}\otimes I)\ket{\Phi_2}
=\ket{a_1}\ket{a_2}$$
Now modify this protocol so that Alice starts with a quantum state
$\ket{a_1a_2}$ and applies $Z^{a_1}X^{a_2}$ to her half of
$\ket{\Phi_2}$ conditioned on her quantum input.  After she sends her
qubit and Bob decodes, they will be left with the state
$\ket{a_1a_2}_A\ket{a_1a_2}_B$.  Thus, 
\be 1\qu + 1\eb \geqslant 2 \cohs \label{eq:SDC-cc}\ee

In fact, any unitary operation capable of classical communication is
also capable of an equal amount of coherent communication,
though in general this only holds asymptotically and for one-way
communication. 

\begin{proposition}\label{prop:unitary-cc}
If $U$ is a bipartite unitary or isometry such that 
$$U + e \,\ebs \geqslant C \cbs\;\asy \label{eq:unitary-cc}$$
then 
$$U + e \,\ebs \geqslant C \cohs \;\asy\,.$$
\end{proposition}

Here we consider $U$ a resource in the sense of \cite{BHLS}, which
considered the asymptotic capacity of nonlocal unitary operations to
generate entanglement and to send cbits.  By appropriate coding (as in
\cite{BS}), we can reduce Proposition~\ref{prop:unitary-cc} to the
following coherent analogue of
Holevo-Schumacher-Westmoreland\cite{HSW} (HSW) coding.

\begin{lemma}[coherent HSW]\label{lemma:HSW}
Given a set of bipartite pure states
$\cS=\{\ket{\psi_x}_{AB}\}_{x=1}^l$, an isometry $U_\psi$ such that
$U_\psi\ket{x}_A=\ket{x}_A\ket{\psi_x}_{AB}$, and an arbitrary
probability distribution $\{p_x\}$ then
$$ U_\psi \geq \chi\cohs + E\,\ebs\;\asy $$
where $E=\sum_x p_x S(\tr_A\psi_x)$ and
$\chi=S\l(\sum_xp_x\tr_A\psi_x\r) - E$.
\end{lemma}

\begin{proof}
One can show\cite{Dev,HSW} that for any $\delta>0,\epsilon>0$ and
every $n$ sufficiently large there exists a code $\cC\subset\cS^n$
with $|\cC|=\exp(n(\chi-\delta))$, a decoding POVM
$\{D_{c}\}_{c\in\cC}$ with error $<\epsilon$ and a type $q$ with
$\|p-q\|_1\leqslant l/n$ such that every codeword $c:=c_1\ldots
c_n\in\cC$ has type $q$ (i.e. $\forall x, |\{c_j=x\}|=nq_x$).  By
error $<\epsilon$, we mean that for any $c\in\cC$,
$\bra{\psi_{c}}(I\otimes D_{c})\ket{\psi_{c}} > 1-\epsilon$.

Using Neumark's theorem\cite{Neu}, Bob can make his decoding POVM into
a unitary operation $U_D$ defined by $U_D\ket{0}\ket{\phi} = \sum_{c}
\ket{c} \sqrt{D_{c}}\ket{\phi}$.  Applying this to his half of a
codeword $\ket{\psi_c}:=\ket{\psi_{c_1}}\cdots\ket{\psi_{c_n}}$ will
yield a state within $\epsilon$ of $\ket{c}\ket{\psi_c}$, since
measurements with nearly certain outcomes cause almost no disturbance.

The communication strategy begins by applying $U_\psi$ to $\ket{c}_A$
to obtain $\ket{c}_A\ket{\psi_c}_{AB}$.  Bob then decodes unitarily
with $U_D$ to yield a state within $\epsilon$ of
$\ket{c}_A\ket{c}_B\ket{\psi_c}_{AB}$.  Since $c$ is of type $q$,
Alice and Bob can coherently permute the states of $\ket{\psi_c}$ to
obtain a state within $\epsilon$ of
$\ket{c}_A\ket{c}_B\ket{\psi_1}^{\otimes nq_1}_{AB}\cdots
\ket{\psi_l}^{\otimes nq_l}_{AB}$.  Then they can apply entanglement
concentration\cite{BBPS} to $\ket{\psi_1}^{\otimes
nq_1}_{AB}\cdots 
\ket{\psi_l}^{\otimes nq_l}_{AB}$ to obtain $\approx nE$ ebits without
disturbing the coherent message $\ket{c}_A\ket{c}_B$.
\end{proof}

There are many cases in which no ancillas are produced, so we do not
need the assumptions of Lemma~\ref{lemma:HSW} that communication is
one-way and occurs in large blocks.  For example, a {\sc cnot} can
transmit one coherent bit from Alice to Bob or one coherent bit from
Bob to Alice.  Given one ebit, a {\sc cnot} can send one coherent bit
in both directions at once using the encoding
\be (H\otimes I)\textsc{cnot}(Y^a\otimes Z^b)\ket{\Phi_2}_{AB} =
\ket{b}_A\ket{a}_B
\label{eq:e-asstd-cnot}
\ee
This can be made coherent by conditioning
the encoding on a quantum register $\ket{a}_A\ket{b}_B$, so that
\be\textsc{cnot} + \eb \geq \coh(\ra) + \coh(\la)
\label{eq:CNOT-power1}\ee
Here $(\ra)$ denotes communication from Alice to Bob and $(\la)$ means
from Bob to Alice.\\

\noindent{\bf Uses of coherent communication }

By discarding her state after sending it, Alice can convert coherent
communication into classical communication, so
$1\coh\geqslant 1\cb$.  Alice can also generate entanglement by
inputting a superposition of messages (as in \cite{BHLS}), so
$1\coh\geqslant 1\eb$.  The true power of coherent
communication comes from performing both tasks---classical communication
and entanglement generation---simultaneously.  This is possible
whenever the classical message sent is random and nearly independent
of the other states at the end of the protocol (as in the
``oblivious'' condition of \cite{obliv-RSP}). 

Teleportation \cite{TP} satisfies these conditions, and indeed a
coherent version has already been proposed in \cite{TP-coh}.  Given an
unknown quantum state $\ket{\psi}_{A}$ and an EPR pair
$\ket{\Phi_2}_{AB}$, Alice begins coherent teleportation not by a Bell
measurement on her two qubits but by unitarily rotating the Bell basis
into the computational basis via a {\sc CNOT} and Hadamard gate.  This
yields the state $\frac{1}{2}\sum_{ij}\ket{ij}_AX^iZ^j\ket{\psi}_B$.
Using two coherent bits, Alice can send Bob a copy of her register to
obtain $\frac{1}{2}\sum_{ij}\ket{ij}_A\ket{ij}_BX^iZ^j\ket{\psi}_B$.
Bob's decoding step can now be made unitary, leaving the state
$\ket{\Phi_2}_{AB}^{\otimes 2}\ket{\psi}_B$.  In terms of resources,
this can be summarized as: $2\cohs + 1\eb \geq 1\qu + 2\ebs$.  Since
the ebit that we start with is returned at the end of the protocol, we
need only use it catalytically: $2\cohs \geq 1\qu + 1\eb \;\cat$.
Combining this relation with \eq{SDC-cc} yields the equality 
\be 2\cohs = 1\qu + 1\eb \;\cat\label{eq:cc-equality}.\ee
Thus teleportation and super-dense coding are reversible so long as
all of the classical communication is left coherent.

Another protocol that can be made coherent is Gottesman's
method\cite{Got} for simulating a distributed CNOT
(i.e. $\ket{x}_A\ket{y}_B\ra\ket{x}_A\ket{x\oplus y}_B$) using one
ebit and one cbit in either direction;
i.e. $1\cb(\ra)+1\cb\mbox{$(\la)$} + 1\eb \geqslant 1\cnot$.  At first
glance, this appears completely irreversible, since a CNOT can be used
to send one cbit forward or backwards, or to create one ebit, but no
more than one of these at a time.

Using coherent bits as inputs, though, allows the recovery of 2 ebits
at the end of the protocol, so $1\coh(\ra)+1\coh(\la)+1\eb \geqslant
1\cnot + 2\ebs$, or using entanglement catalytically,
$1\coh(\ra)+1\coh(\la)\geq 1\cnot+1\eb\,\cat$.  Combined with
\eq{CNOT-power1}, this yields another equality:
$$\cnot+\eb = \coh(\la)+\coh(\ra)\;\cat.$$
Another useful bipartite unitary gate is \textsc{swap}, which is
equivalent to $1\qu(\la)+1\qu(\ra)$, up to catalysis.  Applying
\eq{cc-equality} then yields
$$2\cnot=1\textsc{swap}\;\cat$$
which explains the similar communication and entanglement capacities
for these gates found in \cite{BHLS}.  Previously, the most efficient
methods known to transform between these gates gave $3\cnot\geq
1\textsc{swap} \geq 1\cnot$.\\

\noindent{\bf Remote state preparation}

In remote state preparation (RSP), Alice uses entanglement and
classical communication to
prepare a state of her choice in Bob's lab \cite{RSP}.
Asymptotically, this relation is
$$1\cb+1\eb \geq 1\;\text{remote qubit}\;\asy$$
where ``$n$ remote qubits''
mean the ability of Alice to prepare an arbitrary $2^n$-dimensional
pure state in Bob's lab.  While the entanglement cost of RSP is shown
to be optimal in \cite{RSP}, this lower bound does not necessarily
apply to extended communication protocols that use RSP as a
subroutine.  In particular, using coherent communication
allows all of the entanglement to be recovered at the end of the
protocol, so that asymptotically
\be 1\coh \geq 1\;\text{remote qubit}\;\asy
\label{eq:coherent-rsp}\ee
However, here ``$n$ remote qubits'' means the slightly weaker ability
to prepare $\sqrt{n}$ states, each with $2^{\sqrt{n}}$ dimensions, so
we cannot achieve the full range of remote states possible in
\cite{RSP} without $n$ ebits used as catalysts.

To show \eq{coherent-rsp}, we will need to examine the
protocol in \cite{RSP} more closely.  Alice wishes to transmit a
$d$-dimensional pure state $\psi=\ket{\psi}\bra{\psi}$ to Bob using
the shared state $\ket{\Phi_d}=\frac{1}{\sqrt{d}}\sum_{i=1}^d
\ket{ii}_{AB}$.  For any $\epsilon>0$, choose
$n=\cO(d\log d/\epsilon^2)$.  Ref.~\cite{RSP} proved the
existence of a set of $d\times d$ unitary gates $R_1,\ldots,R_n$ such
that for any $\psi$,
\be 
\l(1-\frac{\epsilon}{2}\r)I \leq
\frac{d}{n}\sum_{k=1}^nR_k\psi R_k^\dag
\leq \l(1+\frac{\epsilon}{2}\r)I 
\label{eq:RSP-lemma}\ee
For $k=1,\ldots,n$ define $A_k=\frac{d}{n(1+\epsilon/2)} (R_k\psi
R_k^\dag)^T$, where the transpose is taken in the Schmidt basis of
$\Phi_d$.  Due to \eq{RSP-lemma}, we have the operator inequalities
$(1-\epsilon) I \leq \sum_k A_k \leq I$.  Thus, if we define
$A_{\text{fail}} = I-\sum_k A_k$, then $A_{\text{fail}}\geq 0$, $\tr
A_{\text{fail}}<d\epsilon$ and ${\bf A}\equiv\{A_1,\ldots, A_n,
A_{\text{fail}}\}$ is a valid POVM.  When ${\bf A} \otimes I$ is
applied to $\Phi_d$, the probability of outcome $A_{\text{fail}}$ is
less than $\epsilon$.  Since for any operator $O$, $(O\otimes I)
\ket{\Phi_d} = (I\otimes O^T)\ket{\Phi_d}$, measurement outcome $k$
leaves Alice and Bob with the state $R_k^*\ket{\bar{\psi}}
\otimes R_k\ket{\psi}$ where
$\psi^T=\ket{\bar{\psi}}\bra{\bar{\psi}}$.

We now show how to apply coherent communication to the above
procedure.  First we apply Neumark's theorem\cite{Neu} to convert {\bf
A} into a unitary operation $\UA$ such that
$$\UA\ket{\varphi}_A\ket{0}_{A'} = \sum_k
\sqrt{A_k}\ket{\varphi}_A\ket{k}_{A'}
+ \sqrt{A_{\text{fail}}}\ket{\varphi}_A\ket{\text{fail}}_{A'}$$
After applying $\UA$, Alice will perform the two-outcome
measurement $\{\sum_k
\ket{k}\bra{k},\ket{\text{fail}}\bra{\text{fail}}\}$ on system $A'$.
The probability of failure is less than $\epsilon$ and upon
success, the resulting state is
$$\frac{1}{\sqrt{n}}\sum_{k=1}^n R_k^*\ket{\bar{\psi}}_A\ket{k}_{A'}
R_k\ket{\psi}_B$$
This can be simplified if Alice applies the unitary operation $\sum_k
\ket{k}\bra{k}_{A'}\otimes (R_k^T)_A$ (i.e. $R_k^T$ to system
$A$ conditioned on the value of system $A'$).  Since $R_k^TR_k^*=I$,
this leaves $A$ in the state $\ket{\bar{\psi}}$, disentangled from the
rest of the system so she can safely discard it. After this, Alice
and Bob share the state $\frac{1}{\sqrt{n}}\sum_{k=1}^n
\ket{k}_{A'}R_k\ket{\psi}_B$.  Alice now uses $\log n\,\cohs$ to
transmit $k$ to Bob coherently, obtaining 
$\frac{1}{\sqrt{n}}\sum_{k=1}^n \ket{k}_{A'}\ket{k}_{B'}
R_k\ket{\psi}_B$.
Bob performs the unitary $\sum_k \ket{k}\bra{k}_{B'} \otimes
(R_k^\dag)_B$ and ends with $\ket{\Phi_n}_{A'B'}\ket{\psi}_B$.

Alice and Bob have gone from $\Phi_d$ to $\Phi_n$, which is a slight
increase in entanglement, though asymptotically insignificant.  Thus $\log n\cohs\geq \log d\text{ remote
qubits}+\log(n/d)\ebs\,\cat$, which implies $1\cc\geq 1\text{ remote
qubit}\,\asy$.  However, for the cost of preparing the inital catalyst
to vanish asymptocally, we need to perform RSP in many separate
blocks, say $\sqrt{n}$ blocks of $\sqrt{n}$ qubits each.
\qed\medskip

\begin{corollary}[RSP capacity of unitary gates]
If $U\geq C \cbs\,\asy$ then $U\geq C\text{ remote qubits }\asy$.
\end{corollary}

\begin{corollary}
{\bf (super-dense coding of quantum states)}
\label{cor:sddc}
$1\qu + 1\eb \geq 2\text{ remote qubits }\asy$
\end{corollary}
Corollary~\ref{cor:sddc} was first proven directly in \cite{SDDC} and
in fact, finding an alternate proof was the original motivation for
this work.

{\em Entangled RSP:}
The states prepared in RSP need not always be completely remote:
\cite{RSP} also showed how Alice can prepare the ensemble
$\cE=\{p_i,\psi_i\}$ of bipartite states using $\chi=S-E$ cbits and
$S$ ebits, where $S=S\l(\sum_i p_i\tr_A\psi_i\r)$ and $E=\sum_i p_i
S(\tr_A \psi_i)$.  Using coherent communication allows
$\chi$ ebits to be recovered, so that
\be \chi\cohs + E\ebs \geq \cE\;\asy.
\label{eq:coh-eRSP}\ee
Here, by $\geq n\cE$, we mean $n$ uses of $U_\psi:\ket{i}_A\ra
\ket{i}_A\ket{\psi_i}_{AB}$ where Alice's input is restricted to the
space spanned by $\ket{i_1\cdots i_n}$ for $p$-typical sequences
$i_1\cdots i_n$.

The proof of \eq{coh-eRSP} is just like the proof of
\eq{coherent-rsp}, but even simpler since the original protocol in
\cite{RSP} already left Alice's state independent of $k$ upon success.
[The same benefits of coherent communication do not apply to the
``low-entanglement'' version of RSP in \cite{RSP}; here entanglement
can only be recovered from the part of the message corresponding to
the measurement.]

Using super-dense coding with \eq{coh-eRSP} allows super-dense coding
of entangled quantum states according to $\chi/2\qus + (E+\chi/2)\ebs
\geq \cE$, a claim for which no direct proof is known.

Coherent RSP of entangled states can also help determine the
communication capacity of bipartite unitary gates as introduced in
\cite{BHLS}.  For any $e\in\mathbb{R}$, define $C_e(U)$ to be the
largest number such that $U + e\,\ebs \geq C_e(U) \cbs\,\asy$.  For
negative values of $e$, $C_e$ corresponds to creating $-e$ ebits per
use of $U$ in addition to communicating $C_e$ cbits; we arbitrarily set
$C_e=-\infty$ when $U\not\geq -e\,\ebs$.
\begin{proposition}\label{prop:Ce-cap}
\ben
C_e(U)&=& \Delta\chi_e(U) \\
&:=& \sup_\cE \l\{ \chi(U\cE)-\chi(\cE)
: E(\cE)-E(U\cE)\leq e\r\}\een
where $\cE=\{p_i,\psi_i\}$ is an ensemble of bipartite states,
$U\cE=\{p_i,U\psi_iU^\dag\}$ and $\chi$ and $E$ are defined as above.
\end{proposition}
Thus the asymptotic capacity equals the largest increase in mutual
information possible with one use of $U$ if the average entanglement
decreases by no more than $e$.  This was proven for $e=\infty$ by
\cite{BHLS}.

{\em  Achieving $C_e(U)\geq\Delta\chi_e(U)$:}

We base our protocol on the one used in Section~4.3 of \cite{BHLS} to show
$C_\infty = \Delta\chi_\infty$.  For any ensemble $\cE$,
\bea
U + \cE &\geq& U\cE \non\\
&\geq & \chi(U\cE)\cohs + E(U\cE)\,\ebs \;\asy
\label{eq:CeU-HSW}\\
&\geq & (\chi(U\cE)-\chi(\cE))\cohs \non\\
&& + (E(U\cE)-E(\cE))\,\ebs + \,\cE
\;\asy\label{eq:CeU-RSP}
\eea
Here \eq{CeU-HSW} used Lemma~\ref{lemma:HSW} and \eq{CeU-RSP} used
\eq{coh-eRSP}.

Now we move the ebits to the left hand side and neglect the catalytic
use of $\cE$ to get
$$U + (E(\cE)-E(U\cE))\,\ebs \geq (\chi(U\cE)-\chi(\cE))\cohs$$
Taking the supremum of the right-hand side over all $\cE$ with
 $E(\cE)-E(U\cE)\leq e$ yields $C_e(U) \geq \Delta\chi_e(U)$.

{\em Proving $C_e(U)\leq\Delta\chi_e(U)$:}

Without loss of generality, we can assume Alice and Bob defer all
measurements until the end of the protocol, so at all points we work
with pure state ensembles.  Since $\chi(\cE)$ and $E(\cE)$ are
invariant under local unitaries and non-increasing under the final
measurement, we need only consider how they are modified by $U$.
Thus, $n$ uses of $U$ and $ne$ ebits can increase the mutual
information by at most $n\sum_j p_j \Delta\chi_{e_j}$ for some $\sum_j
p_j=1$ and $\sum_j p_je_j \leq e$.  Our bound will follow from proving
that $\Delta\chi_e$ is a concave function of $e$.

To prove concavity, consider a probability distribution $\{p_j\}$,
a set of ensembles $\cE_j$ and the new ensemble $\cE=\{ p_j,
\ket{jj}_{AB} \otimes \cE_j\}$.  Taking the simultaneous supremum of
$\chi(U\cE_j)-\chi(\cE_j)$ over
each $\cE_j$ subject to
$E(\cE_j)-E(U\cE_j)\leq e_j$ completes the proof.  \qed\medskip

{\em Quantum capacities of unitary gates} We can also consider the
ability of unitary gates to send quantum information.  Define $Q_e$ to
be the largest number such that $U+e\,\ebs \geq Q_e\qus$.  Then from
\eq{cc-equality}, one can obtain
\be Q_e = \frac{1}{2} C_{e + Q_e} \ee
where $C_e$ can be determined from Proposition~\ref{prop:Ce-cap}.

\noindent{\bf Conclusions}

Coherent communication offers a new way of looking at
quantum information resources in which irreversible transformations
occur only when coherence is discarded and not just because we are
transforming incomparable resources.  Whenever classical
communication is used in a quantum protocol, either as input or
output, and the classical message is nearly independent of the
other quantum states, there may be gains from making the communication
coherent.

The case of noisy coherent communication remains to be fully solved
and it would be interesting to find the coherent capacity of
a noisy channel, or more generally to find their $C_e$ and $Q_e$
tradeoff curves.  Some preliminary results in these directions have
been obtained\cite{family}.

\begin{acknowledgments}
My thanks go to C.H.~Bennett, P.~Hayden, D.~Leung, J.~Smolin,
A.~Winter and especially I.~Devetak for helping me come up with and
refine the ideas in this paper.  I am also grateful for the
hospitality of the Caltech IQI, IBM T.J. Watson research center and
the ERATO center.  My funding is from the NSA and ARDA under ARO contract
DAAD19-01-1-06.

\end{acknowledgments}

\end{document}